\documentclass[twoside,twocolumn,9pt]{article}

\usepackage{extsizes}
\usepackage[super,sort&compress,comma]{natbib} 
\usepackage[left=1.5cm, right=1.5cm, top=1.785cm, bottom=2.0cm]{geometry}
\usepackage{balance}
\usepackage{widetext}
\usepackage{times,mathptmx}
\usepackage{sectsty}
\usepackage{graphicx} 
\usepackage{lastpage}
\usepackage[format=plain,justification=raggedright,singlelinecheck=false,font={stretch=1.125,small,sf},labelfont=bf,labelsep=space]{caption}
\usepackage{float}
\usepackage{fancyhdr}
\usepackage{fnpos}
\usepackage[english]{babel}
\usepackage{array}
\usepackage{charter}
\usepackage[T1]{fontenc}
\usepackage[usenames,dvipsnames]{xcolor}
\usepackage{setspace}
\usepackage[compact]{titlesec}
\usepackage{ulem}

\usepackage{epstopdf}

\definecolor{cream}{RGB}{222,217,201}

\begin{document}

\pagestyle{fancy}
\thispagestyle{plain}


\makeFNbottom
\makeatletter
\renewcommand\LARGE{\@setfontsize\LARGE{15pt}{17}}
\renewcommand\Large{\@setfontsize\Large{12pt}{14}}
\renewcommand\large{\@setfontsize\large{10pt}{12}}
\renewcommand\footnotesize{\@setfontsize\footnotesize{7pt}{10}}
\makeatother

\renewcommand{\thefootnote}{\fnsymbol{footnote}}
\renewcommand\footnoterule{\vspace*{1pt}%
\color{cream}\hrule width 3.5in height 0.4pt \color{black}\vspace*{5pt}} 
\setcounter{secnumdepth}{5}

\makeatletter 
\renewcommand\@biblabel[1]{#1}            
\renewcommand\@makefntext[1]%
{\noindent\makebox[0pt][r]{\@thefnmark\,}#1}
\makeatother 
\renewcommand{\figurename}{\small{Fig.}~}
\sectionfont{\sffamily\Large}
\subsectionfont{\normalsize}
\subsubsectionfont{\bf}
\setstretch{1.125} 
\setlength{\skip\footins}{0.8cm}
\setlength{\footnotesep}{0.25cm}
\setlength{\jot}{10pt}
\titlespacing*{\section}{0pt}{4pt}{4pt}
\titlespacing*{\subsection}{0pt}{15pt}{1pt}


\makeatletter 
\newlength{\figrulesep} 
\setlength{\figrulesep}{0.5\textfloatsep} 

\newcommand{\topfigrule}{\vspace*{-1pt}%
\noindent{\color{cream}\rule[-\figrulesep]{\columnwidth}{1.5pt}} }

\newcommand{\botfigrule}{\vspace*{-2pt}%
\noindent{\color{cream}\rule[\figrulesep]{\columnwidth}{1.5pt}} }

\newcommand{\dblfigrule}{\vspace*{-1pt}%
\noindent{\color{cream}\rule[-\figrulesep]{\textwidth}{1.5pt}} }

\makeatother

\title{Structuring polymer gels via catalytic reactions}
\twocolumn[
  \begin{@twocolumnfalse}
\vspace{3cm}
\sffamily
\begin{tabular}{m{4.5cm} p{13.5cm} }

  & 
\noindent\LARGE{\textbf{Structuring polymer gels via catalytic reactions}} \\
\vspace{0.3cm} & \vspace{0.3cm} \\

 & \noindent\large{Virginie Hugouvieux\textit{$^{a}$} and Walter Kob\textit{$^{b}$}} \\
\vspace{0.3cm} & \vspace{0.3cm} \\
& 
\noindent\normalsize{We use computer simulations to investigate how a catalytic reaction in a polymer sol can induce the formation of a polymer gel. To this aim we consider a solution of homopolymers in which freely-diffusing catalysts convert the originally repulsive A monomers into attractive B ones. We find that at low temperatures this reaction transforms the polymer solution into a physical gel that has a remarkably regular mesostructure in the form of a cluster phase, absent in the usual homopolymer gels obtained by a quench in temperature. We investigate how this microstructuring depends on catalyst concentration, temperature, and polymer density and show that the dynamics for its formation can be understood in a semi-quantitative manner using the interaction potentials between the particles as input. The structuring of the copolymers and the AB sequences resulting from the reactions can be discussed in the context of the phase behaviour of correlated random copolymers. The location of the spinodal line as found in our simulations is consistent with analytical predictions. Finally, we show that the observed structuring depends not only on the chemical distribution of the A and B monomers but also on the mode of formation of this distribution. }

\end{tabular}
 \end{@twocolumnfalse} \vspace{0.6cm}

]


\renewcommand*\rmdefault{bch}\normalfont\upshape
\rmfamily
\section*{}
\vspace{-1cm}


\footnotetext{\textit{$^{a}$~SPO, INRA, Montpellier SupAgro, University of Montpellier, 34060 Montpellier, France. E-mail: virginie.hugouvieux@inra.fr}}
\footnotetext{\textit{$^{b}$~Laboratoire Charles Coulomb, UMR 5221, University of Montpellier and CNRS, 34095 Montpellier, France. E-mail: walter.kob@umontpellier.fr}}



\section*{Introduction}

Many properties of polymer gels that are formed by physical or chemical cross-linking are well understood~\cite{rubinstein2003polymers} and thus these materials are used in a multitude of applications~\cite{bohidar,djabourov,food_gels,drug_delivery,tissue_engineering,tissue_hydrogel,webber2016}. But polymeric gels are also found in living organisms~\cite{biological_hydrogels,biological_gels,Cosgrove} and in these cases the cross-linking is often related to the presence of enzymes, i.e.~biological catalysts, which can trigger the formation or breaking of covalent or non-covalent crosslinks. This unusual type of mechanism for forming or degrading gels, or more generally materials, is, e.g., at work in the case of enzyme-induced formation and/or degradation of gels made of proteins~\cite{Giraudier, Lairez}, peptides~\cite{Ulijn}, or polysaccharides~\cite{Payne}, with the prominent example of plant cell walls in which pectins form a physical gel due to the action of an enzyme~\cite{Cosgrove}. In this latter case a freely moving catalyst converts the initially repulsive monomers of the polymers into attractive ones, making that with time the polymers increasingly attract each other and as a consequence slowly transform the polymer sol into a gel. Despite their relevance for many living organisms the structural properties and the dynamics of formation of such gels have so far been explored very little. 

Earlier studies on phase-separating systems undergoing chemical reactions~\cite{glotzer1994,glotzer1995,krishnan2015,tanaka1992,kyu1996,tran1996} have evoked the possibility of controlling their steady-state morphology by tuning the interplay between reaction rate and phase separation but so far no specific attempts have been made to follow up this idea in a more quantitative manner. The goal of the present work is to investigate the microstructure  of polymeric gels formed via catalytic reactions as well as its evolution with time and thus to advance our understanding on what type of structures are formed and how they depend on time. Obtaining this insight will be an important step forward for using these systems in a variety of material science applications. Moreover we will see that the chemical correlations along the polymer chains generated by catalytic reactions are surprisingly similar to the ones found in random copolymers studied in previous theoretical~\cite{fredrickson1991PRL,fredrickson1992Macromol,sung2005integral,sung2005JCP,mao2016flexibility} and simulation works~\cite{houdayer2002EPL,houdayer2004Macromol,gavrilov2011ChemPhysLett,slimani2013Macromol}. Therefore the results on the structure and dynamics of our catalyst-induced gels are also useful to progress our understanding on the properties of random copolymer melts. 

\section*{Model and simulation details}

In our simulations the polymers are modelled as bead-spring chains consisting of two types of monomers, A and B, that have the same size $\sigma$ and mass $m$. At the beginning of the simulation we start with a solution of homopolymers constituted of A monomers. These monomers interact with each other through a purely repulsive Weeks-Chandler-Andersen (WCA) potential~\cite{WCA} obtained by truncating a Lennard-Jones potential, $V_{\rm LJ}(r)=4 \varepsilon \left[ (\sigma/r)^{12} - (\sigma/r)^6 \right]$, at the distance $r=2^{1/6}\sigma$ and shifting it to zero by adding $\varepsilon$. (Here $r$ is the distance between two monomers, $\sigma$ characterizes their size, and $\varepsilon$ is the depth of the potential well.) In addition to this hard-core potential, the monomers are connected by a finite extensible nonlinear elastic (FENE) potential~\cite{FENE} of the form $V_{\rm FENE}(r)= -0.5 k r_0^2 \ln (1-(r/r_0)^2) $ with $k=30 \varepsilon/\sigma^2$ and $r_0=1.5 \sigma$. In the following we will express length and energy in units of  $\sigma$ and $\epsilon$, respectively, time in units of $\tau=\sqrt{m\sigma^2/\varepsilon}$, and temperature $T$ in units of $\varepsilon$, setting the Boltzmann constant equal to 1.0.

Also present in the melt are the catalysts which we model as soft sphere particles of size $\sigma_c=2\sigma$ and mass $m_c=5m$. These particles interact with the A monomers also by a WCA potential with an interaction radius given by the mean of $\sigma$ and $\sigma_c$. The non-attractive interaction between monomers and catalysts corresponds to the fact that in the context of the plant cell walls the catalysts do not have any specific affinity for their substrate. (However, in Nature other cases exist as well and it would certainly be interesting to study them as well.) Once the mixture of polymers and catalysts is thoroughly equilibrated at a given temperature $T$, we allow the catalysts, which so far were inert particles, to transform an A monomer into a B monomer. These B monomers interact with the A monomers and the catalysts in the same manner as the A monomers do. However, the interaction between two B monomers is attractive and given by a Lennard-Jones potential that is truncated and shifted at 2.5$\sigma$. 

A catalyst can induce a transformation of an A monomer into a B monomer whenever its distance from the monomer is below $d_{\rm reac} = \frac{1}{2}(\sigma+\sigma_c)-0.07$. This reaction occurs only with a probability of 10\%, since in real systems catalyst activity is less than 100\% due to thermal noise and the orientation of the catalytic molecule~\cite{fersht1999,biocatalysts}. The bare barrier $\Delta_{\rm AC}$ for the reaction of an A monomer with a catalyst is the value of the WCA potential at $d_{\rm reac}$ which is found to be 2.77, a value that we will see below is important for the dynamics. Whenever the A to B conversion occurs, a linear transition (duration 100 simulation steps) from the WCA to the LJ potential is done in which the local energy of the converted monomer and its neighbours of type B is conserved by rescaling their kinetic energies (in addition to the thermostat).

Simulations are carried out in the $NVT$ ensemble using the LAMMPS software~\cite{LAMMPS}, to which we have added the ability to perform reactions between the catalysts and A monomers. The simulated systems consist of 408 chains with 100 monomers, thus $N_m=40800$ monomers, and a catalyst-to-monomer number ratio $N_C / N_m$ ranging from $0.005$ to $0.05$, but, if not stated explicitly otherwise, all the data in the text are for $N_C/N_m=0.012$. The simulated monomer densities $\rho_m=N_m/L^3$ (with $L$ the size of the simulation box) range from $0.2$ to $0.6$ corresponding to monomer volume fractions in the range $[0.1,0.3]$, relevant for the gelation of polymers of this length. Trajectories are generated using the velocity-Verlet integrator with a timestep $h=0.003$. 

\begin{figure}[!th]
\centering
\includegraphics[width=0.9\linewidth]{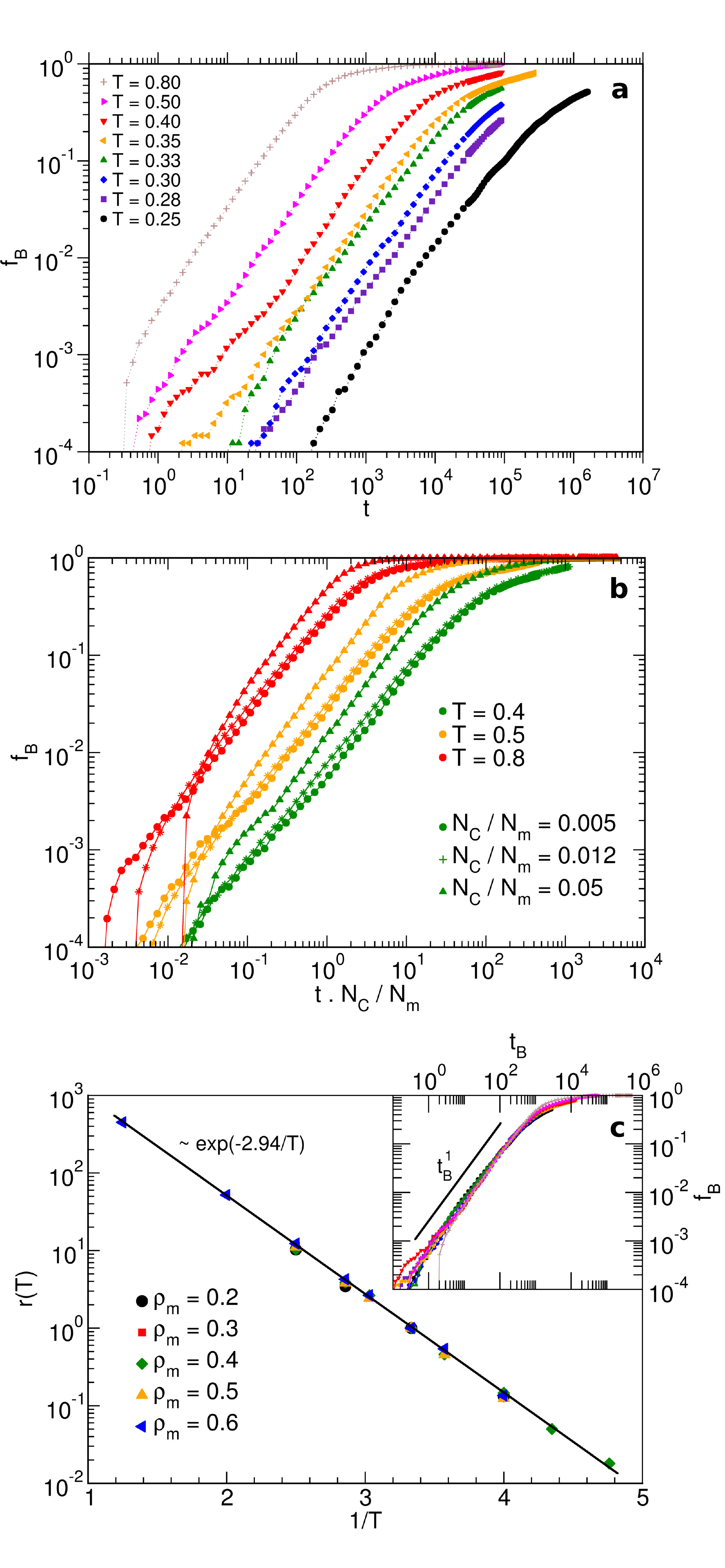}
\caption{(Colour online) 
(a) Fraction of B monomers as a function of time $t$ for different temperatures and $\rho_m=0.6$. (b)  Influence of the catalyst concentration on the A$\rightarrow$B reaction rate. For catalyst-to-monomer ratios $N_C / N_m$ below $\approx$0.012, the transformation rate scales linearly with $N_C / N_m$ and thus the curves for the different concentrations fall on top of each other. Note that same colour corresponds to same $T$ and same symbol to same $N_C / N_m$. To improve the statistics of the results the curves for  $N_C / N_m = 0.012$ have been averaged at short times over 8 independent samples. (c) Temperature and density dependence of the scaling factor $r(T)$. Symbols: $r(T)$ used to define the time scale $t_B$. The scaling factor shows a simple Arrhenius dependence on temperature with an activation energy that is independent of the polymer density $\rho_m$ (solid line). Inset: Same data as (a) but now as a function of the reduced time $t_B= t \cdot N_C / N_m \cdot r_0 \cdot \exp(-2.94/T)$ (where $r_0$ is a constant) showing that this scaling leads to collapse of the data onto a master curve.}
\label{fig:nb_B_arrhenius}
\end{figure}

\section*{Catalytic conversion of the monomers}
Due to the transformation of the A monomers into B monomers by the catalysts, the number of B monomers $N_B$ in the system increases with time, see Fig.~\ref{fig:nb_B_arrhenius}a. This figure shows that the fraction of B monomers, $f_B=N_B/N_m$, first evolves linearly with time $t$ for almost three decades, indicating that in this time window the motion of the catalyst particles allows them to constantly find A monomers they can convert. A more detailed analysis of this motion shows that it is diffusive, i.e. the mean squared displacement is linear in time (not shown).

In Fig.~\ref{fig:nb_B_arrhenius}b we show the influence of the number density of the catalyst particles on the reaction kinetics for different values of $T$. Note that time is scaled by the catalyst-to-monomer ratio $N_C/N_m$. We recognize that the curves for $N_C/N_m \le 0.012$ nicely superimpose, demonstrating that at small concentration the transformation rate of A to B is directly proportional to $N_C/N_m$. For higher concentrations, $N_C/N_m = 0.05$, the reaction kinetics is slightly faster, indicating that collective effects start to affect the transformation dynamics. We also mention that this acceleration is accompanied by an increase of the {\it total} volume fraction, which is somewhat surprising since normally glass-forming systems show a slowing down of their dynamics with increasing particle density~\cite{binderkob2011}. We therefore speculate that this acceleration is related to the presence of collective effects in the motion of the catalysts.

Figure~\ref{fig:nb_B_arrhenius}a shows that at long times the concentration of B monomers saturates since only few A monomers are left and the catalyst particles cannot reach them any more because the former are trapped in dense B regions. We also recognize that at short times the linear $t-$dependence is present for all temperatures considered and that it slows down if $T$ is decreased. To determine the influence of temperature on the kinetics of the A$\rightarrow $B reaction for a given value of the monomer density $\rho_m$, we have rescaled the curves of Fig.~\ref{fig:nb_B_arrhenius}a onto a master curve by plotting them as a function of $t_B = t \cdot N_C / N_m \cdot r(T)$, where $r(T)$ is a scaling factor used to collapse the data. (Here the factor $N_C/N_m$ takes into account the above mentioned linear dependence of the reaction rate on $N_C$.) We find that the $T-$dependence of $r(T)$ is given by an Arrhenius law with an activation energy $E_B = 2.94$ (see main graph of Fig.~\ref{fig:nb_B_arrhenius}c) that is independent of $\rho_m$ and is very close to $\Delta_{\rm AC}= 2.77$, the bare potential energy barrier between an A monomer and a catalyst. Hence at short and intermediate $t$ the time scale for the A $\rightarrow$ B conversion is just given by the time needed to overcome this  barrier which is independent of temperature and density. The inset of Fig.~\ref{fig:nb_B_arrhenius}c shows $f_B$ as a function of $t_B=t \cdot N_C / N_m \cdot r_0 \cdot \exp(-2.94/T)$, where $r_0$ is a constant determined from the main panel of Fig.~\ref{fig:nb_B_arrhenius}c, and we see that this representation leads indeed to a master curve.

\begin{figure}[t] 
\centering
\includegraphics[width=0.95\linewidth]{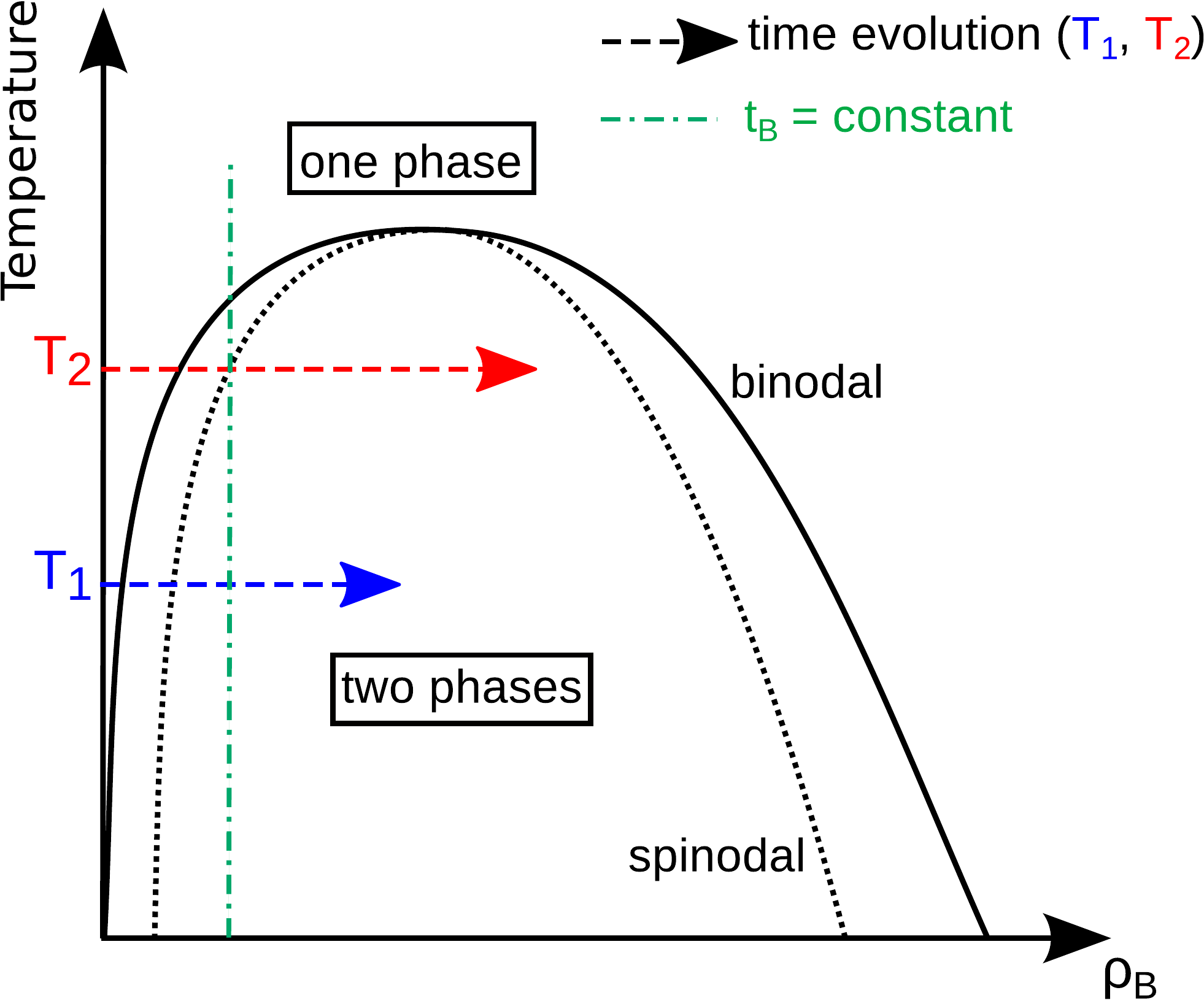}
\caption{(Colour online) Schematic phase diagram illustrating the evolution of the system with time. The horizontal axis is the concentration of the attractive B monomers. The binodal and spinodal lines are for a homo-polymer system that has only  attractive B-monomers. The dashed arrows for temperatures $T_1$ and $T_2$ depict the expected time evolution of the system as $\rho_B$ increases due to the catalytic reactions. The vertical dash-dotted line indicates $t_B$=const., i.e.~a constant thermodynamic driving force.}\label{fig:phase_diagram}
\end{figure}

\section*{Evolution of the structure}
By definition of $t_B$, systems at different $T$ but with the same value of $t_B$ and $\rho_m$ have the same fraction of B monomers $f_B$ and hence the same number density $\rho_B$ of B monomers. Since at low $T$ the presence of the attractive B monomers will give rise to a phase separation, their concentration can be expected to be directly related to a (time-dependent) driving force for this thermodynamic instability. Thus an increasing $\rho_B$ will drive the system into the coexistence region and consequently induce the phase separation. This mechanism is sketched in Fig.~\ref{fig:phase_diagram} where we show in the $\rho_B-T$ plane the coexistence and spinodal lines for a system in which all the monomers are of type B (full and dotted lines, respectively). The {\it real} system will move in this phase diagram along the dashed arrows that indicate the evolution of the concentration of B monomers with increasing time. As $\rho_B$ increases the system will enter the metastable region between the binodal and spinodal lines where nucleation and growth may occur and then reach the unstable region (below the spinodal) where it performs phase separation.

\begin{figure}[!tb]
\centering
\includegraphics[width=\linewidth]{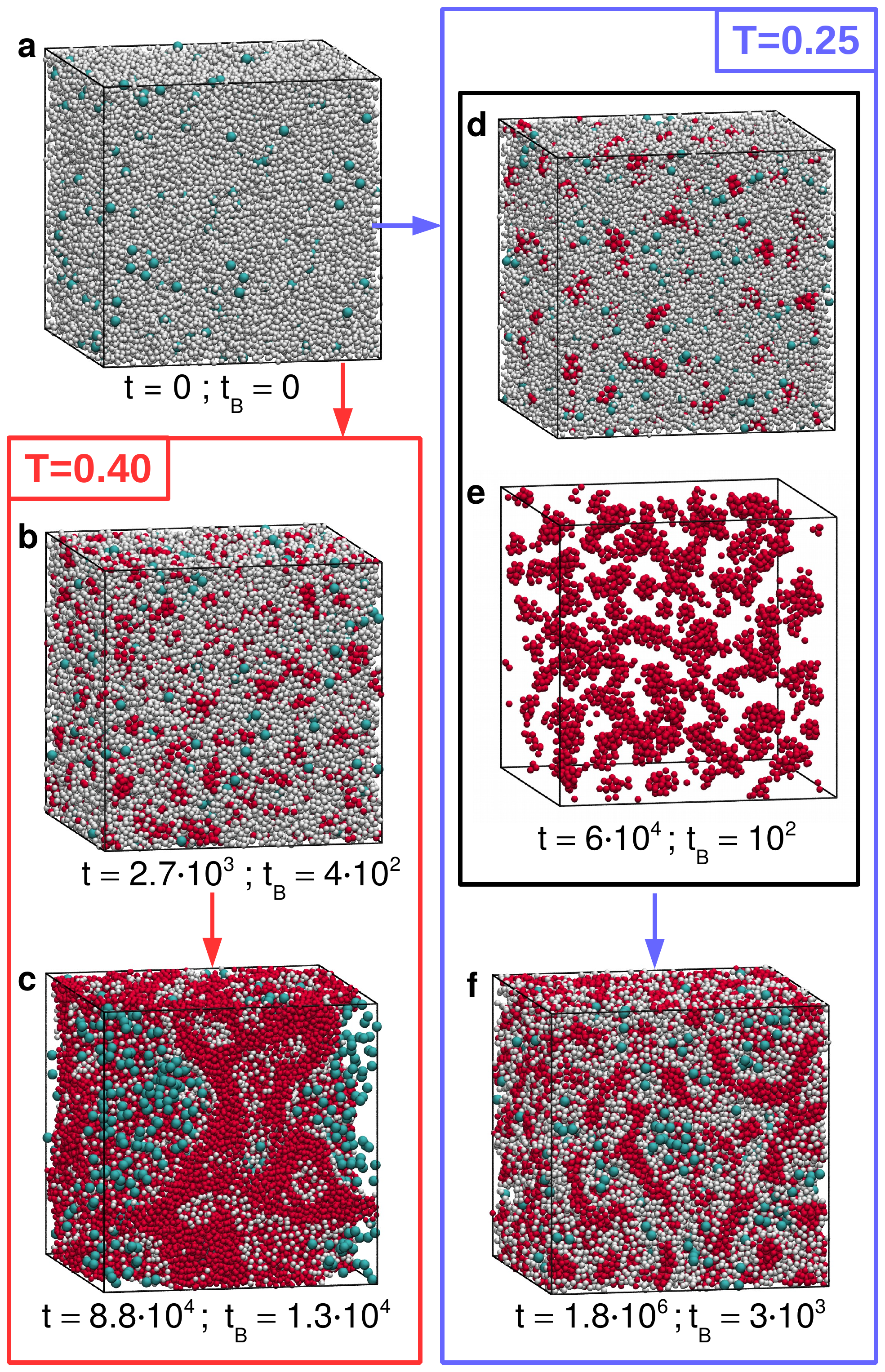}
\caption{(Colour online) Evolution of the structure at $\rho_m=0.6$. 
(a)~$t=0$: Homo-polymer solution that contains catalysts (A monomers in white, catalysts in blue).  With increasing time, A monomers are converted into B monomers (red) which aggregate and form clusters, panels (b) and (d). At high $T$ and long times the macroscopic phase separation starts, panel (c),  whereas at low $T$ the system forms a gel, panel (f). Panel (e) is the same configuration as in panel (d), but now only the B monomers are shown so that the clustering becomes more visible.} \label{fig:snapshot_vs_tb}
\end{figure}

The snapshots of the system, Fig.~\ref{fig:snapshot_vs_tb}, allow to get a qualitative understanding of its evolution with time. They indicate that the structure does indeed transform in a manner that is qualitatively similar to a spinodal decomposition: At high $T$ and long times it forms domains that are polymer rich or polymer poor (Fig.~\ref{fig:snapshot_vs_tb}c) whereas at low $T$ it forms a gel (Fig.~\ref{fig:snapshot_vs_tb}f). 
From panel c of Fig.~\ref{fig:snapshot_vs_tb} we recognize that when the system starts to phase separate on the mesoscopic length scale, most of the catalysts are expelled from the polymer rich phase. This effect is due to entropic reasons since the increasing concentration of the attractive B monomers leads to a tightening of the polymer network and hence to the creation of large empty cavities. As a consequence the catalysts can gain entropy by avoiding the dense polymer network and instead move into the cavities. A side effect of this partial phase separation is the reduction of the local density of the catalysts in the polymer rich phase and hence to a decrease in the speed of the A$\rightarrow$B conversion (as seen in Fig.~\ref{fig:nb_B_arrhenius}a for long times).
Most remarkable is the observation that at intermediate times the system shows an unexpected mesostructure that contains clusters of B monomers (see Fig.~\ref{fig:snapshot_vs_tb}b, \ref{fig:snapshot_vs_tb}d, \ref{fig:snapshot_vs_tb}e) and in the following we will discuss the $t-$dependence of this structural evolution in more detail.

\section*{Time scale for cluster formation}
Although a constant $\rho_B$ implies a constant driving force, the {\it response} to this force and hence the evolution of the system can be expected to depend on temperature. This is certainly true for the formation of the {\it local} structure since the local energy is expected to be the relevant quantity for this formation. To take into account this $T-$dependent response we make the Ansatz that the relevant time scale which determines the structure of the system at intermediate times is given by $t_S=t_B \cdot s(T,\rho_m)$, i.e.~the product of the time scale for the driving force and a factor $s(T,\rho_m)$ that characterizes the dynamic response of the system. We determine this factor $s(T,\rho_m)$ by requiring that systems with the same $t_S$ and $\rho_m$ but at different $T$ have the same structure. This iso-structure time $t_S$ thus allows to compare for short and intermediate times the properties of systems at different temperatures. We emphasize that it is not evident at all that such a time $t_S$ really exists, i.e.~that the evolution of the system can be described by just one internal time scale $t_S$. To test this hypothesis in practice we use $s(T,\rho_m)$ as free parameter to superimpose $S_B(q)$, the static structure factor of the B-particles, where $q$ is the wave-vector.

\begin{figure}
\centering
\includegraphics[width=0.99\linewidth]{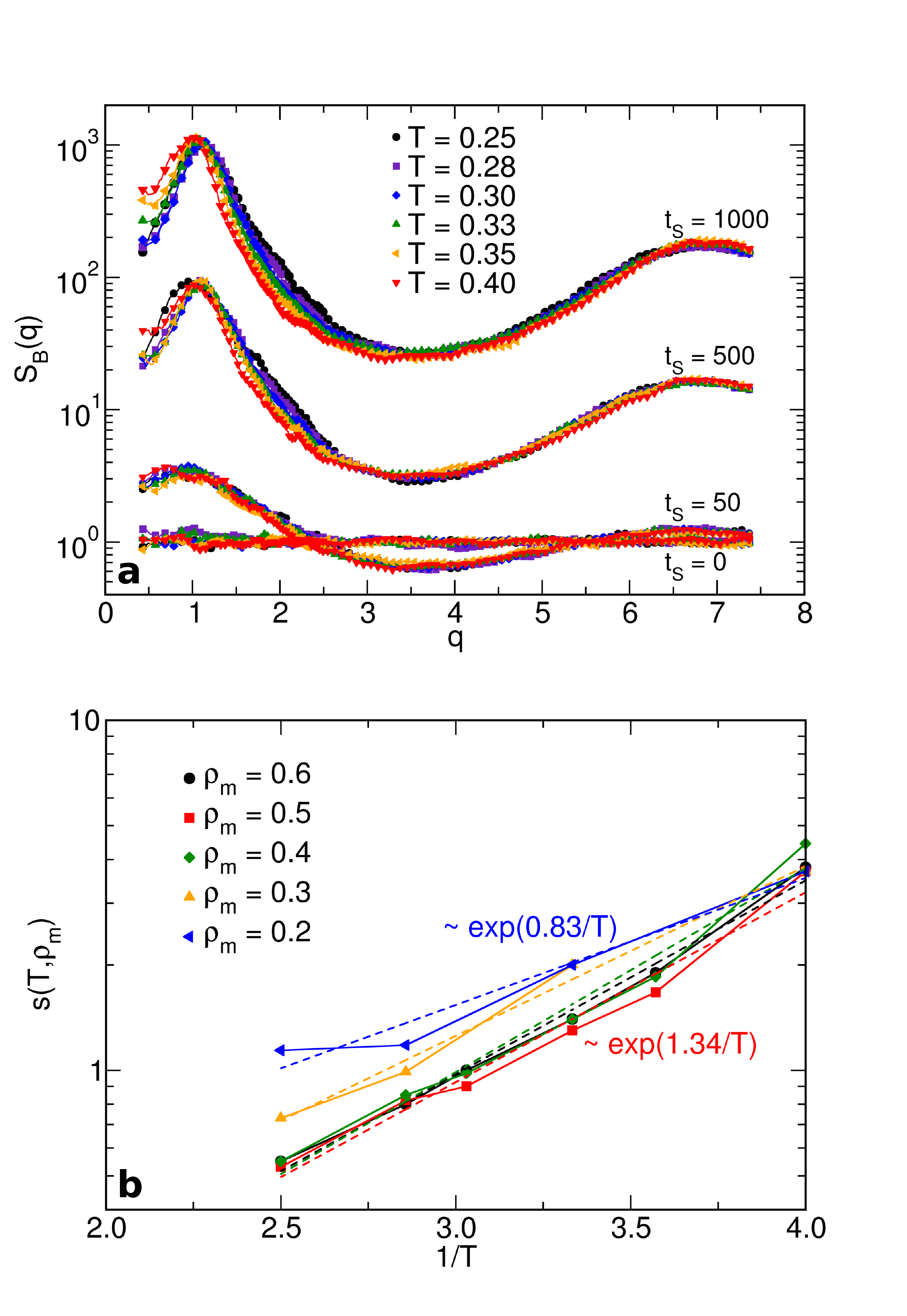}
\caption{(Colour online)
(a)~Structure factor $S_B(q)$ at $\rho_m=0.6$ for different times $t_S$ and temperatures. For clarity $S_B(q)$ is multiplied by 10 and 100 for $t_S=500$ and $t_S=1000$, respectively. (b)~Temperature and density dependence of $s(T,\rho_m)$. Symbols: $T-$dependence of the scaling factor $s(T,\rho_m)$ for the different values of $\rho_m$. Within the accuracy of the data this dependence is given by an Arrhenius law (dashed lines). The activation energy is $\approx 1.34$ for $\rho_m \ge 0.4$ and decreases with $\rho_m$ for $\rho_m < 0.4$.} \label{fig:max_Sbq_ts}
\end{figure}

\begin{figure}
\includegraphics[width=\linewidth]
{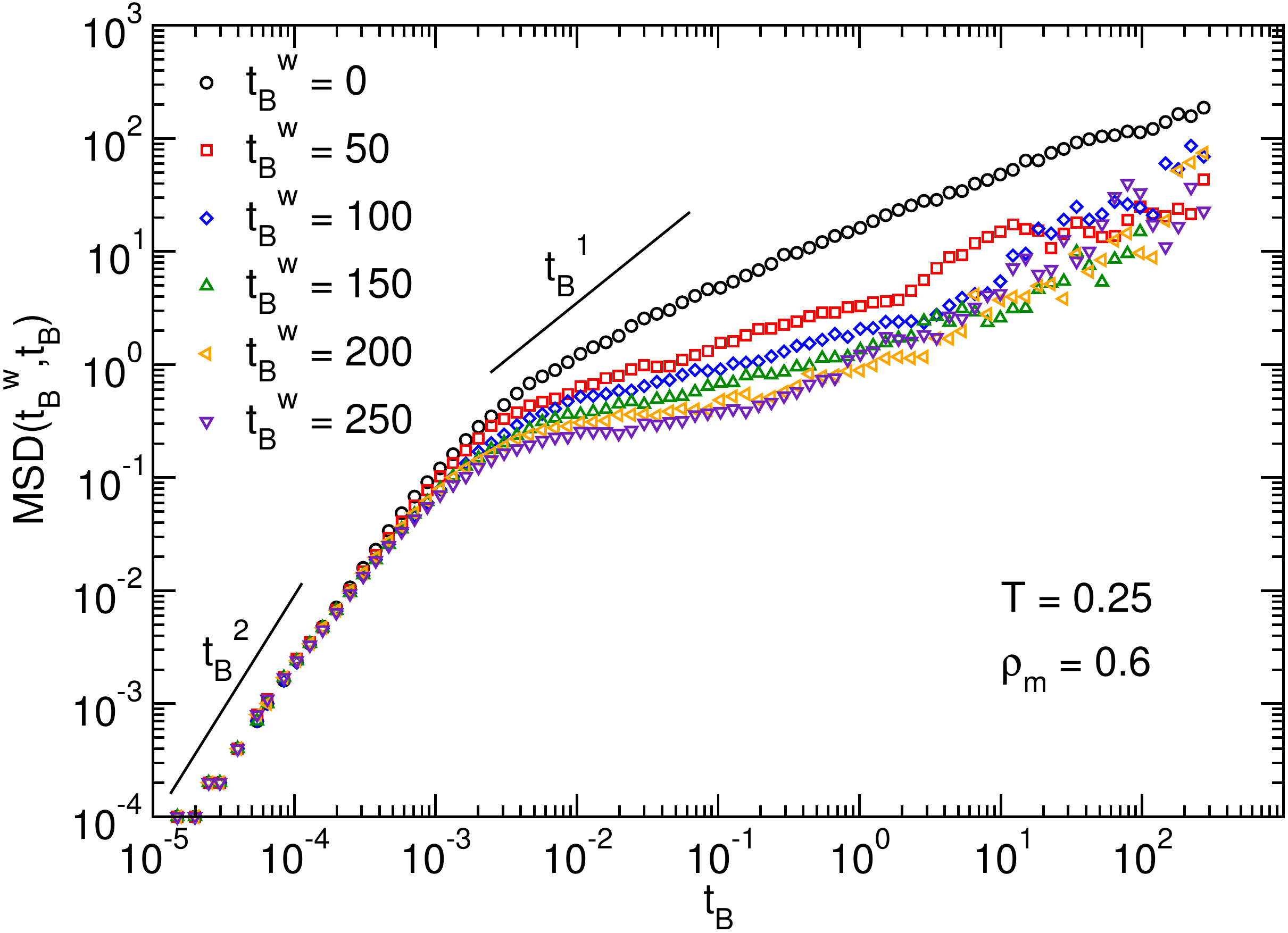}
\caption{(Colour online) Mean square displacements of the centres of mass of the polymers as a function of $t_B$, for $T=0.25$ and $\rho_m=0.6$. The different curves correspond to different waiting times $t_B^w$.}
\label{fig:msd}
\end{figure}

Figure~\ref{fig:max_Sbq_ts}a demonstrates that it is indeed possible to find a scaling factor $s(T,\rho_m)$ that leads to a master curve for the $S_B(q)$ at the different temperatures. If $t_S$ is small there are only very few B monomers and hence $S_B(q)$ is basically flat. With increasing time one finds a peak at $q \approx 7$, corresponding to the nearest neighbour distance between two monomers. At the same time we see a marked peak at $q_p \approx 1.0$. This value is basically independent of time, indicating that the peak is {\it not} related to a standard coarsening process that shows a growing length scale~\cite{Tanaka}, but is instead directly linked to the clusters of B particles seen in the snapshots. (We have also carried out simulations of the same reacting system but without any connectivity between the monomers and found that such a system shows the normal coarsening dynamics and no peak at $q_p\approx 1.0$. Therefore this peak is indeed related to the fact that we consider polymers.) Note that the peak is quite high, indicating that these clusters have a well-defined distance from each other and that the length scale $2\pi/q_p\approx 6$ corresponds indeed to the distance between neighbouring clusters seen in the snapshots. We also mention that the position of this peak is not related to the radius of gyration of the polymers since $q_p$ is the same for $N_m=100$ and $N_m=200$ (data not shown). At present it is thus not evident what polymer intrinsic length scale selects the wave-vector $q_p$. It is remarkable that this rather complex $q-$dependence of $S_B(q)$ is present at all temperatures (but at different times) and that the structure factors can be superimposed with very good accuracy by choosing just one scaling factor $s(T,\rho_m)$. This implies that on the time scale considered the relaxation dynamics of the system can indeed be parametrized by a single internal variable, the iso-structure time $t_S$. Note that the good superposition of the curves starts to deteriorate at small $q$ once $t_S$ has reached $O(10^3)$ in that the peak becomes wider, see Fig.~\ref{fig:max_Sbq_ts}a. This shows that on these time scales it is no longer possible to define a system-intrinsic time scale $t_S$.

\begin{figure}
\centering
\includegraphics[width=0.99\linewidth]{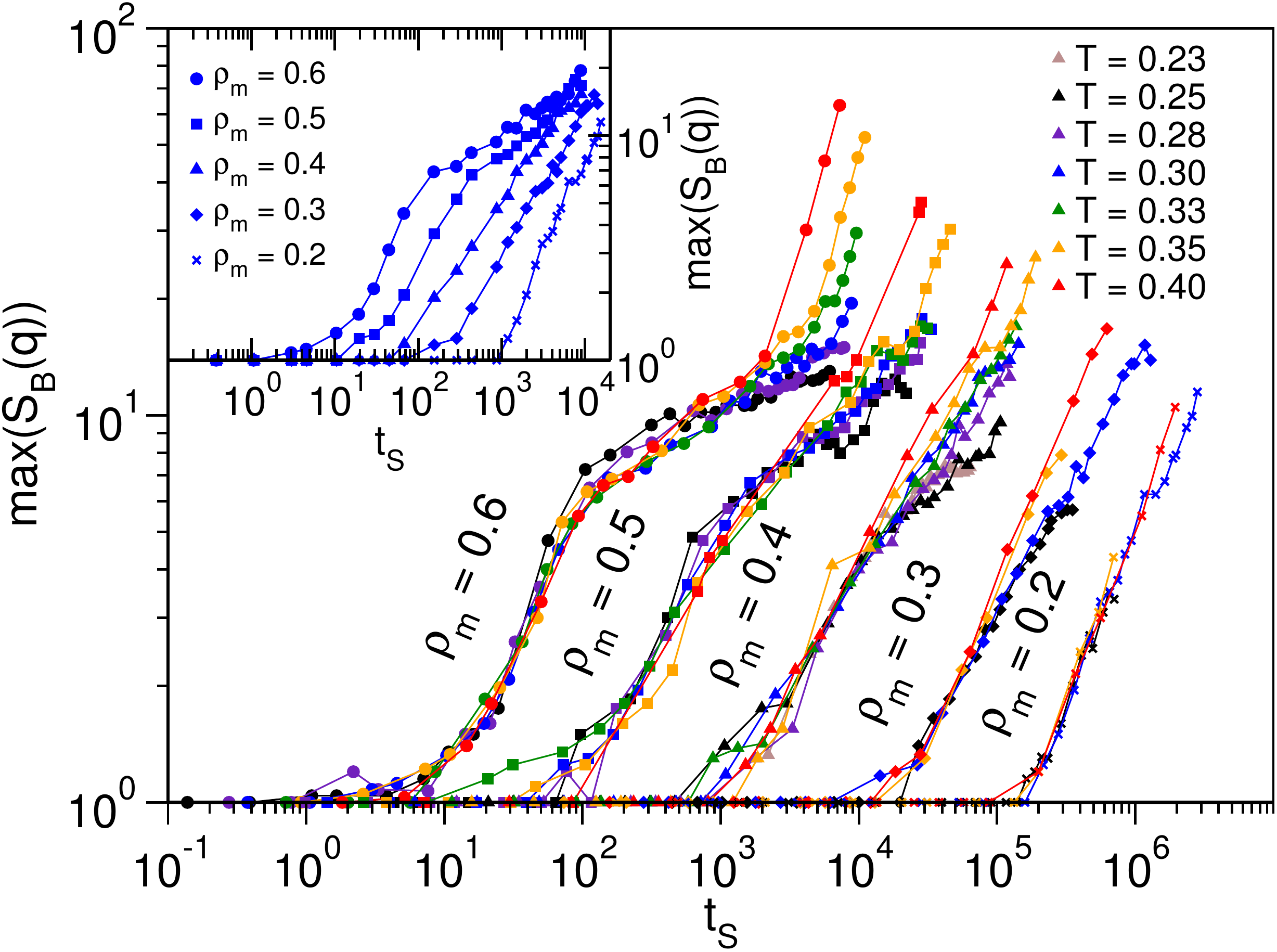}
\caption{(Colour online) Evolution of the maximum of $S_B(q)$ as a function of $t_S$ for different values of $T$ and $\rho_m$. Same type of symbols for same $\rho_m$, same colour for same $T$. For the sake of clarity the curves are shifted to the right by a factor of 4, 16, 64, and 128 for $\rho_m=0.5$, $0.4$, $0.3$, and $0.2$, respectively. Inset: Same quantity at $T=0.30$ and different values of $\rho_m$. } \label{fig:max_Sbq_ts_vs_rhom}
\end{figure}

\begin{figure}
\centering
\includegraphics[width=\linewidth]{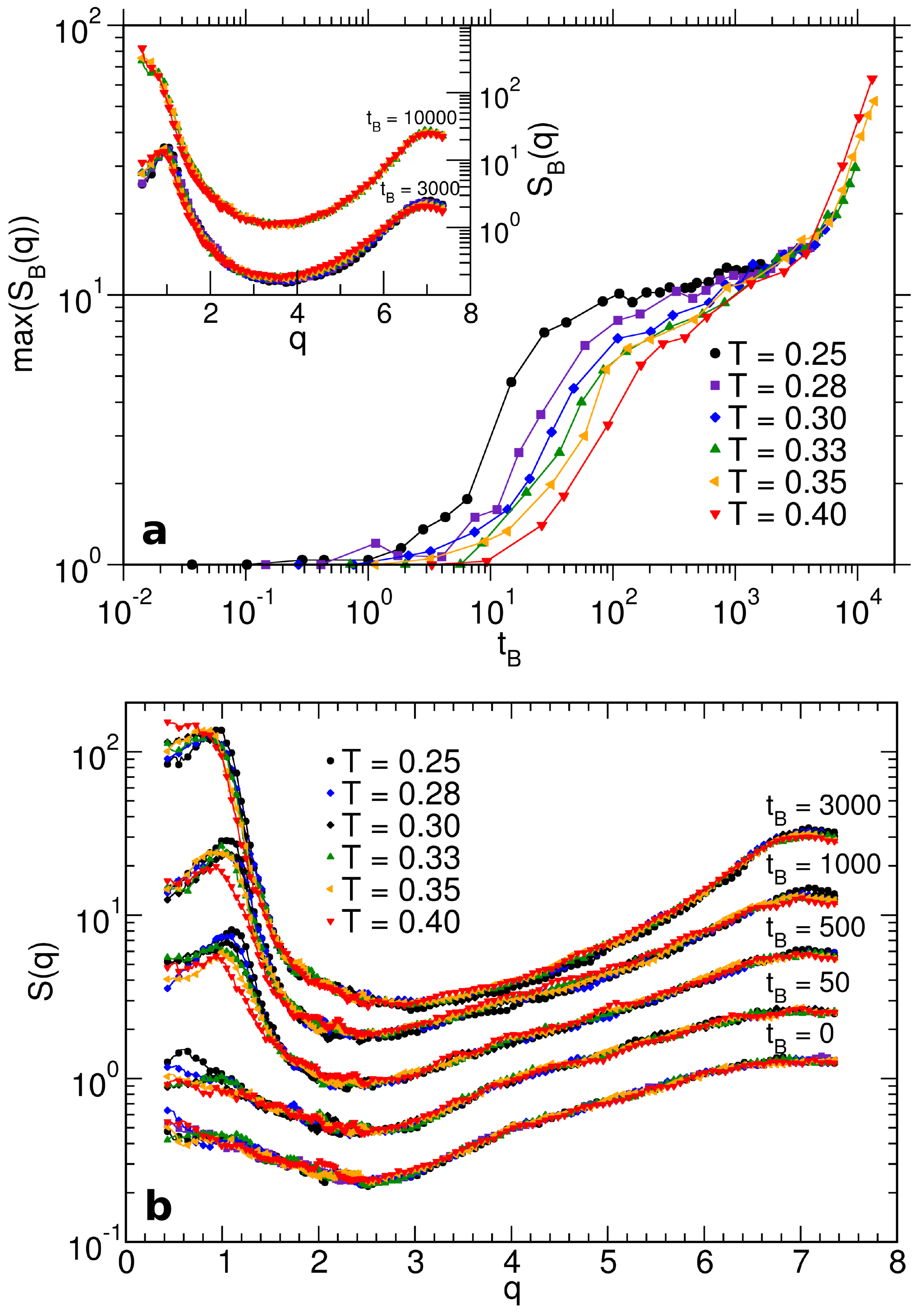}
\caption{(Colour online) (a)~Evolution of the maximum in the static structure factor $S_B(q)$ as a function of $t_B$ for different values of $T$. Inset: $S_B(q)$ for different values of $T$ and $t_B$. Note that for clarity $S_B(q)$ for $t_B=10000$ is multiplied by 10. (b)~Total structure factor $S(q)$ at $\rho_m=0.6$ and different values of $T$ and $t_B$. Note that for clarity $S(q)$ is multiplied by 2, 4, 8, and 12 for $t_B=50$, $500$, $1000$, and $3000$, respectively.} \label{fig:max_Sbq_tb}
\end{figure}

We have found (see Fig.~\ref{fig:max_Sbq_ts}b) that $s(T,\rho_m)$ follows an Arrhenius law with an activation energy $E_s \approx 1.3$ for $\rho_m \ge 0.4$, which is roughly comparable with the well depth $\varepsilon=1.0$ of the Lennard-Jones potential governing the attraction between B monomers. Hence we can conclude that the dynamic response of the system that leads to the evolution of its structure is directly related to the attraction between B monomers.


While Fig.~\ref{fig:snapshot_vs_tb}c demonstrates that at high $T$ the system does make a complete phase separation, we find that at low $T$ the system remains in the disordered state even at long times, i.e.~it forms a gel, Fig.~\ref{fig:snapshot_vs_tb}f. That this system is indeed a gel can be recognized by probing the mean squared displacements of the polymers defined as 

\begin{equation}
MSD(t_B^w,t) = \frac{1}{N}\sum_i^N \left< (r_i(t_B^w+t) - r_i(t_B^w)) ^2\right> \quad ,
\end{equation}

\noindent
where $t_B^w$ is the (waiting) time since the start of the reaction. (Note that since we are studying an out-of-equilibrium system, the $MSD$ will depend not only on the time difference but also on the starting time of the measurement.) Figure~\ref{fig:msd} shows the time dependence of the $MSD$ for different values of the waiting time $t_B^w$. For small $t_B^w$ the ballistic motion seen at short times, i.e~$MSD \propto (t_B)^2$, crosses over directly to a marked sub-diffusive dynamics at long times, i.e~$MSD \propto t_B^\alpha$ with $\alpha\approx 0.6$. With increasing $t_B^w$ we see that at intermediate times the $MSD$ shows a plateau, i.e.~the hallmark of a caging dynamics of the particles~\cite{binderkob2011,kob1995}. We also note that the height of this plateau is significantly larger than the one found in dense glass-forming systems~\cite{kob1995} which shows that the cages are relatively large, i.e~the system is indeed a gel in which the particles relax slowly but can undergo fluctuations with relatively large amplitudes.

To quantify the evolution of the structure we have determined the height of the peak at $q_p \approx 1.0$, and show its $t-$dependence in Fig.~\ref{fig:max_Sbq_ts_vs_rhom}. For the density that we have discussed so far ($\rho_m=0.6$, leftmost set of data) the curves for the different temperatures fall at short and intermediate times nicely on a master curve, thus demonstrating that it is indeed possible to define a single time scale $t_S$ that describes the structural relaxation leading to the formation of the clusters. The observed master curve increases first quickly before it crosses over to a much slower (logarithmic) time dependence. For even longer times the curves bend again upwards and $t_S$ is no more the relevant timescale, a result that is coherent with the observation made in the context of Fig.~\ref{fig:max_Sbq_ts}a that for $t_S\approx O(10^3)$ the curves at small $q$ do no longer superimpose. Note that the extent of the quasi plateau at intermediate times depends strongly on $T$ and thus, if $T$ is sufficiently low, the final step of the phase separation can be moved to very large times, thus allowing to form a stable gel.

The $t$-dependence that we have just discussed is for a high density of polymers. If $\rho_m$ is reduced the initial increase of the peak at $q_p$ is delayed because the B monomers are on average farther apart and the plateau at intermediate times is less pronounced (see Inset of Fig.~\ref{fig:max_Sbq_ts_vs_rhom}). For low densities the plateau seems to disappear completely, at least in the temperature range that we show here. (But we mention that for $\rho_m=0.4$ we have carried out simulations at even lower $T$, $T=0.23$, and found that the plateau is present even at this density, see main panel of Fig.~\ref{fig:max_Sbq_ts_vs_rhom}.) However, for all cases we can define a $\rho_m-$dependent scaling factor $s(T,\rho_m)$ that allows to superimpose the peak height at $q_p$ for short and intermediate times (main panel of Fig.~\ref{fig:max_Sbq_ts_vs_rhom}). Hence we can define an iso-structure time $t_S=t_B \cdot s(T,\rho_m)$ that allows to compare for short and intermediate times the structure of systems at different temperatures. We have tested that for fixed $\rho_m$ systems with the same $s(T,\rho_m)$ do indeed have the same $S_B(q)$ for all $q$, in agreement with the results shown in~Fig.~\ref{fig:max_Sbq_ts}a for the case $\rho_m=0.6$. Figure~\ref{fig:max_Sbq_ts}b shows that for all considered values of $\rho_m$ the scaling factor $s(T,\rho_m)$ has a $T-$dependence that is given by an Arrhenius law. For $\rho_m \ge 0.4$ the activation energy is a constant and given by $E_s \approx 1.34$ whereas it depends on $\rho_m$ for $\rho_m \le 0.3$. Since all of these factors have the same Arrhenius dependence and the same activation energy if $\rho_m$ is not too small, we can conclude that in this case the growth of the clusters is completely governed by the B-B interaction.

\section*{Time scale for coarsening}
It is also instructive to study the time-dependence of the structure as a function of the time scale $t_B$, i.e.~as a function of the concentration of B monomers, since this time scale is related to the thermodynamic driving force. The $t_B-$dependence of the height of the peak at $q_p\approx 1$ is shown in the main graph of Fig.~\ref{fig:max_Sbq_tb}a from which we recognize that at short and intermediate times this peak grows faster if $T$ is decreased. This behaviour is reasonable since at low $T$ the entropic fluctuations are reduced and hence the system can reach more easily an energetically favourable structure, i.e.~form the clusters. At long times and high $T$ the system makes a complete phase separation (see Fig.~\ref{fig:snapshot_vs_tb}c), in agreement with the observation that $S_B(q)$ grows at low $q$ a peak (Inset of Fig.~\ref{fig:max_Sbq_tb}a), as usual in coarsening systems~\cite{Tanaka}. Note that the corresponding $t-$dependence shown in the main graph of Fig.~\ref{fig:max_Sbq_tb}a is not related to a growth of the peak at $q\approx 1.0$ but instead to the growth of the structure factor at small wave-vectors due to the coarsening that sets in at late times (see Fig.~\ref{fig:max_Sbq_tb}a, Inset). When the system starts coarsening on the mesoscopic scale, the curves for different $T$'s superimpose which indicates that the coarsening process is indeed completely governed by the concentration of the B monomers, i.e. the effective attractive interaction between the polymers (see also Fig.~\ref{fig:max_Sbq_tb}a, Inset). This is supported by the fact that the {\it total} structure factors at large fixed $t_B$ but different temperatures superimpose for all wave-vectors (see Fig.~\ref{fig:max_Sbq_tb}b for $t_B=3000$).
Thus we conclude that the phase separation is driven by the fraction of B monomers at long times, but that the structuring of the system at short and intermediate times is governed by the time scale $t_S$, which makes that there is {\it no} good superposition of the curves in Fig.~\ref{fig:max_Sbq_tb}b (at small $q$), but a good one in Fig.~\ref{fig:max_Sbq_ts}a.

\section*{Internal structure of the chains}
\begin{figure}
\includegraphics[width=\linewidth]{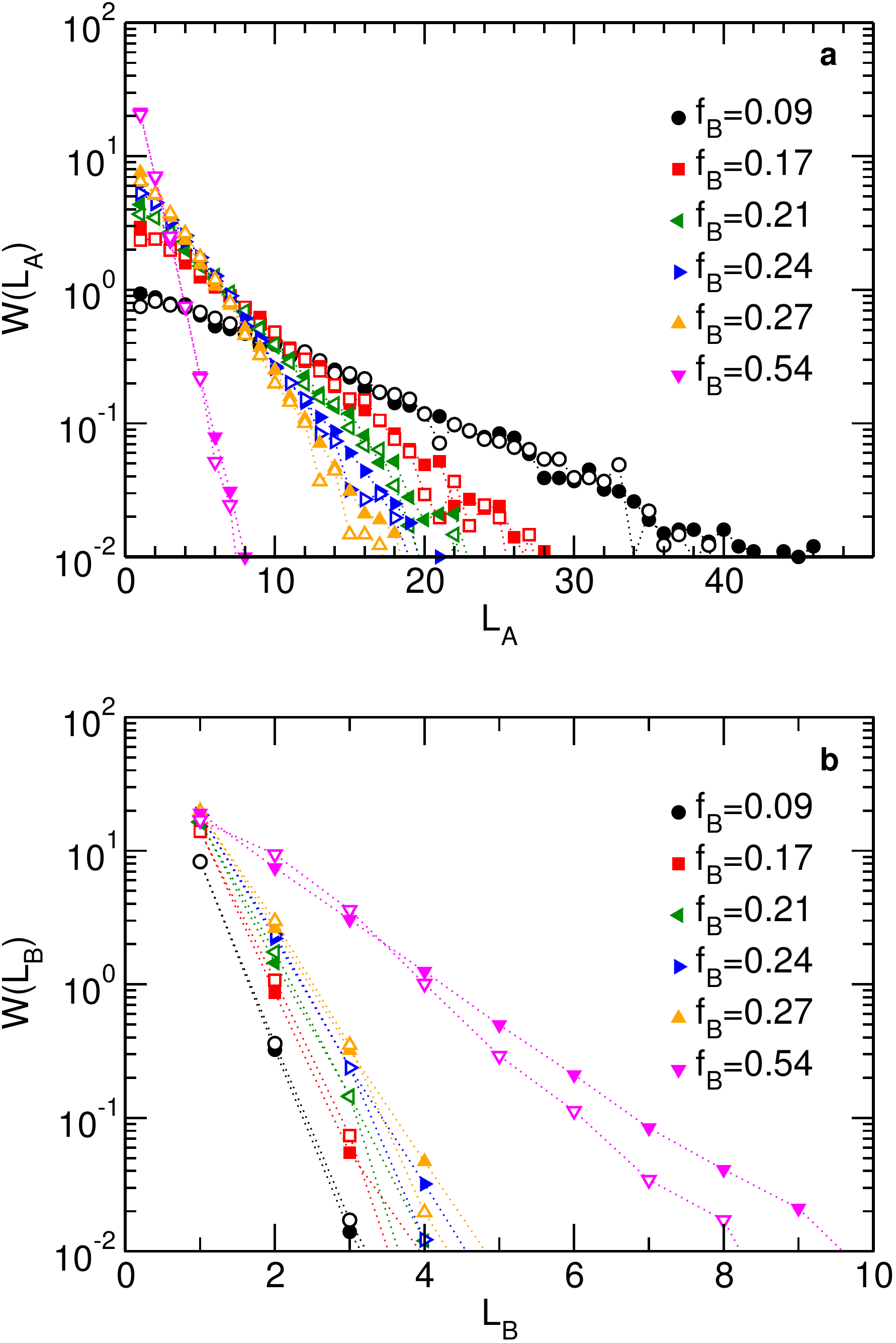}
\caption{(Colour online) Distribution of the lengths of the A blocks, $L_A$ (panel a) and of the B blocks, $L_B$ (panel b). Filled symbols are for the copolymers resulting from the catalytic reaction and open symbols are for the randomly generated A-B copolymers. Curves with the same colour correspond to the same values of  $f_B$, $p_{AA}$ and $p_{BB}$. $T=0.25$. Note that $f_B$ increases with increasing simulation (or reaction) time.}
\label{fig:block_distrib}
\end{figure}

We have shown in the previous sections that the action of the catalysts on the polymers induces the progressive conversion of A monomers into B monomers, which in turn triggers at low temperatures the formation of a cluster phase with a liquid-like organization. In the present section we discuss the internal structure of the chains ({\it i.e.} the distribution of A and B monomers along the chains) and how it can be related to the global structure of the gel.

To characterize the internal structure of the polymers we have determined the distribution of the lengths of blocks of pure A monomers, $W(L_A)$, and blocks of pure B monomers, $W(L_B)$, within a chain, see Fig.~\ref{fig:block_distrib}a and \ref{fig:block_distrib}b, filled symbols. In the context of random copolymer melts, i.e.~if the A and B monomers are randomly distributed on the chain with, respectively, probability $f_A$ and $f_B=1-f_A$, one often specifies also $p_{AA}$, the conditional probability that a A monomer is immediately followed by another A monomer, and $p_{BB}$, the analogous probability for the B monomers, i.e.~one considers not only completely random copolymers but takes into account also the first nearest neighbour correlation. From these probabilities it is then possible to calculate $W(L_A)$ and $W(L_B)$~\cite{fredrickson1991PRL,fredrickson1992Macromol,sung2005integral}. The quantity $\lambda=p_{AA}+p_{BB}-1$ determines whether copolymers are completely random, $\lambda=0$,  while $\lambda=1$ and $\lambda=-1$ correspond to the case of homopolymers and alternating A-B copolymers, respectively.

In the following we will show that the internal structure of the polymers as obtained from the catalytic reaction can indeed be very well described by the one of random block copolymers. For this we have generated a large number of random A-B copolymers with B monomer fraction $f_B$ and probabilities $p_{AA}$ and $p_{BB}$. The values $f_B$, $p_{AA}$, and $p_{BB}$ were obtained from the simulations of the catalyst-induced copolymers at different values of $T$ and at different stages of the reactions, {\it i.e.} different values of $f_B$ in the simulations.

For $T=0.25$ the block length distributions of the catalytically (filled symbols) and randomly generated (open symbols) copolymers are shown in Fig.~\ref{fig:block_distrib} for different values of $f_B$. For small and intermediate $f_B$ the two ways of generating copolymers give rise to very similar block length distributions and only for large $f_B$ noticeable differences are seen in that the blocks from the catalytic reactions are a bit longer than the ones from the random copolymers. Thus we can conclude that at short and intermediate times the catalytic reaction does indeed give rise to an internal structure of the polymers that is very similar to the one of random copolymers. We also note that random copolymers generated by imposing only $f_B$ but not $p_{AA}$ and $p_{BB}$ have a block length distribution that differs significantly from the one of the catalyst-induced copolymers. Hence the sequences of A-B monomers in the catalyst-induced copolymers can be considered as random but  correlated, and the structuring of these copolymers can be discussed in the theoretical framework developed for random copolymers~\cite{fredrickson1991PRL,fredrickson1992Macromol,sung2005integral,sung2005JCP}. 

\begin{figure}
\includegraphics[width=\linewidth]
{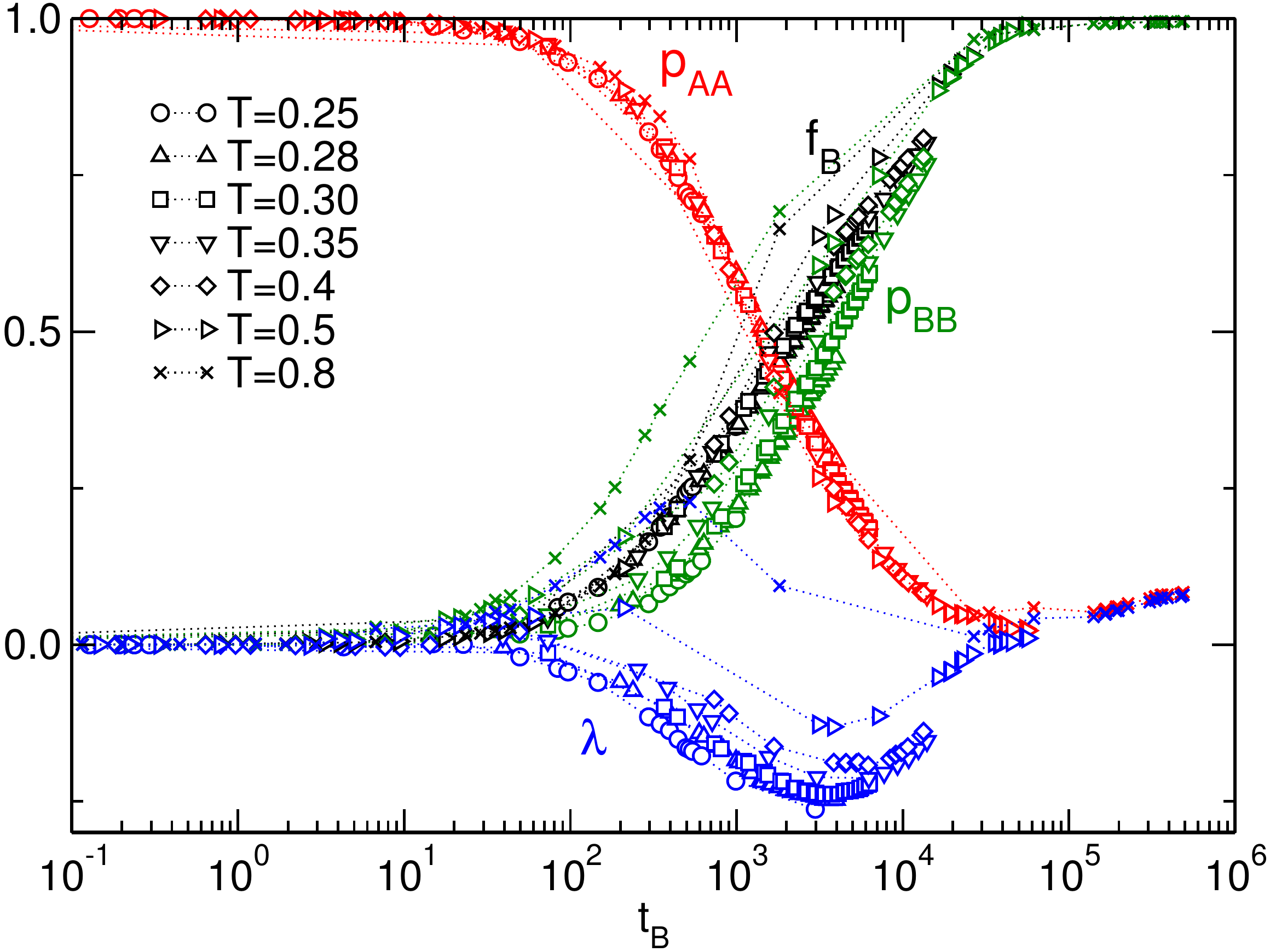}
\caption{(Colour online) Time evolution of the chemical correlations for different values of $T$: Fraction of B monomers $f_B$ (black), conditional probabilities $p_{AA}$ (red) and $p_{BB}$ (green), and $\lambda$ (blue).}
\label{fig:Sq_reac_vs_T_v2}
\end{figure}

The time dependence of the parameters $f_B$, $p_{AA}$, $p_{BB}$, and $\lambda$ is shown in Fig.~\ref{fig:Sq_reac_vs_T_v2} for different values of $T$. Note that we plot the data as a function of $t_B$, since this is the time scale that is directly related to the time dependence of $f_B$. We see that for intermediate and low temperature each set of data falls on a master curve. Hence we can conclude that $t_B$ is indeed the relevant variable that determines the internal structure of the chain. This internal structure determines in turn the relative arrangements of the chains and hence $t_B$ is the relevant time scale for the structure of the {\it system}, in agreement with our results shown in Fig.~\ref{fig:max_Sbq_tb}b.

We see that the $t_B-$dependence of $p_{BB}$ is quite similar to the one of $f_B$ (which is also shown in Fig.~\ref{fig:nb_B_arrhenius}) and that the probability $p_{AA}$ decays on roughly the same time scale as $p_{BB}$ is growing. However, these two time-dependences are not exactly the same as can be seen from the $t_B-$dependence of $\lambda$ which shows at intermediate times a local minimum, i.e.~$p_{AA}$ is decaying faster than $p_{BB}$ is growing. Since close to the minimum $\lambda$ is negative we can conclude that in this time window the arrangement of the A and B monomers is such that on average one has smaller blocks of purely A (or purely B) monomers than expected from a completely random chain that has B monomers with probability $f_B$, or put otherwise, the monomer sequence is alternating more rapidly than in a random chain. The reason for this enhanced alternation is likely related to the fact that two B monomers attract each other and hence form a local domain that has a higher than average density. For entropic reasons the catalytic particles will thus have the tendency to avoid these crowded regions and will instead be more concentrated in regions where there are mainly A monomers. These latter regions will be more likely to have chains that contain relatively large blocks of A monomers and the presence of the catalysts will make that these large blocks are cut into smaller pieces. Thus effectively the catalysts will be more likely to break up a larger block than a smaller one, thus leading to a proliferation of rather short blocks, i.e.~a $\lambda$ that is negative.

As mentioned above, there have been earlier theoretical studies of the  phase diagram of random copolymers. Using a mean-field approach Fredrickson and Milner have determined the phase diagram of random copolymer melts as a function of the chemical correlations within the chains characterized by $\lambda$ \cite{fredrickson1991PRL}. They predicted the existence of a spinodal line, i.e.~macrophase separation when $T$ is lowered below a certain threshold that depends on $\lambda$ or a microphase separation. Using the PRISM integral equation theory, Sung and Yethiraj refined these results by including hard-core interactions~\cite{sung2005integral,sung2005JCP} and found that the details of the phase diagram depend quite strongly on the details of the closure approximations~\cite{sung2005integral}. A recent field-theoretical study considered the influence of chain rigidity on the onset of phase separation and on the typical size of the corresponding domains~\cite{mao2016flexibility}. These analytical predictions have been followed up by computer simulations of various polymeric systems. Using Monte Carlo simulations Houdayer and M\"uller showed that a coarse grained random copolymer melt undergoes a macroscopic phase separation followed by a disordered microemulsion-like phase if $T$ is decreased at large values of $\lambda$, while no phase separation is obtained for $\lambda$ smaller than a critical value~\cite{houdayer2002EPL,houdayer2004Macromol}. Subsequently Gavrilov \textit{et al.} showed that random copolymers can form lamellar phases in the super strong segregation regime\cite{gavrilov2011ChemPhysLett} and more  recently Slimani and coworkers\cite{slimani2013Macromol} used molecular-dynamics simulations to show that random copolymers (with fixed $f_B$) can form microdomains which typical size does not depend on temperature.

The fact that the internal structure of our polymers is very similar to the one of random copolymers, see Fig.~\ref{fig:block_distrib}, suggests that the meso-structure that we have found to form during the catalytic reaction, see Fig.~\ref{fig:max_Sbq_ts}, is related to the thermodynamic instability predicted to be present in random copolymers. Although it is rather difficult to test this connection in a quantitative manner, we will see that at least qualitatively the theory matches well our simulation data.

\begin{figure}
\includegraphics[width=\linewidth]{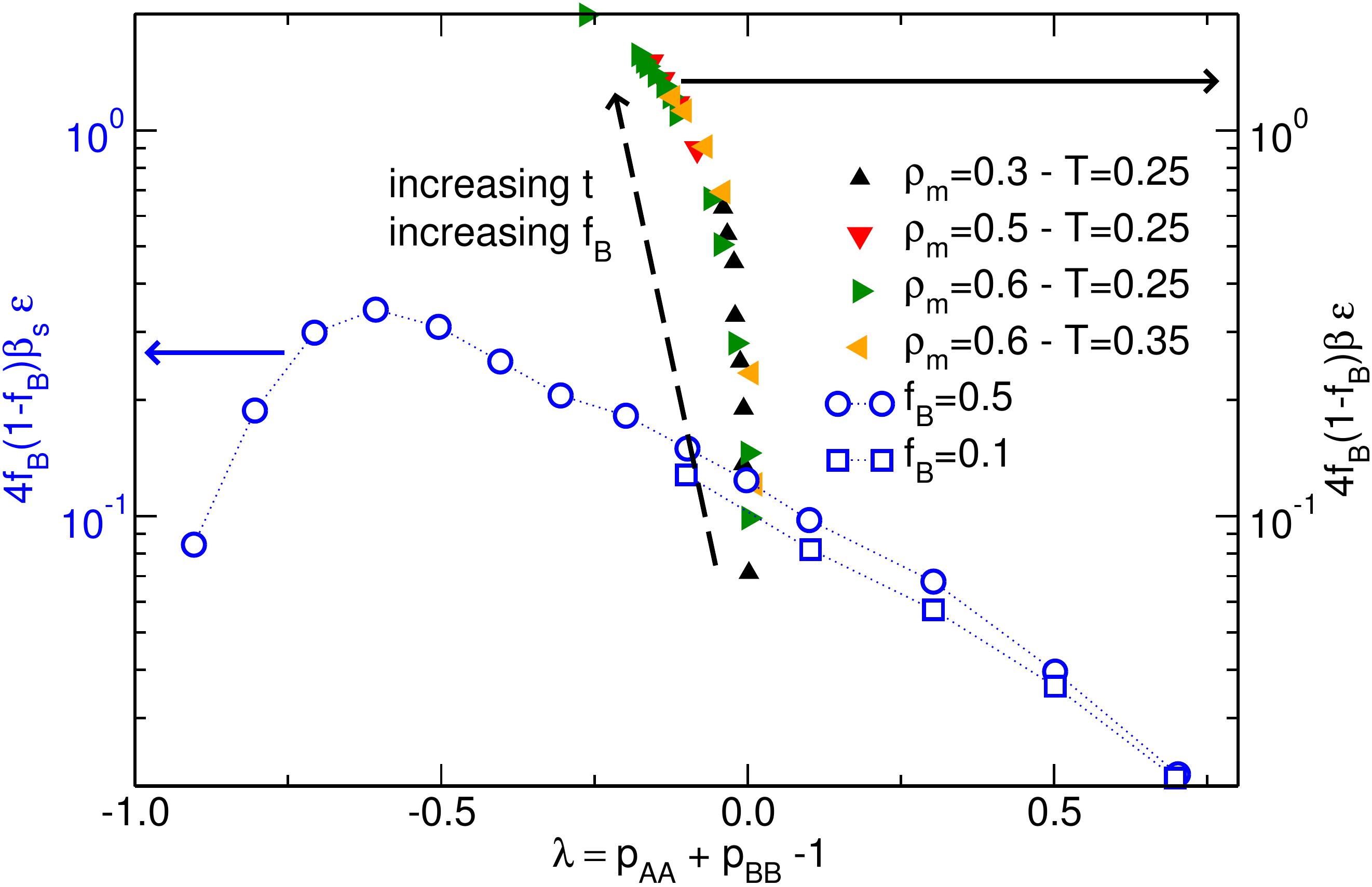}
\caption{(Colour online) Blue open symbols and left scale: Spinodal lines for a system of random copolymers as determined from the PRISM integral equation theory for two values of $f_B$~\cite{sung2005integral}. Full symbols and right scale: Points obtained from the simulation at different values of $T$ and $\rho_m$. Note that the left scale of the ordinate depends on $\beta_s$ which is the inverse spinodal temperature while the right scale depends on $\beta$, the inverse temperature.}
\label{fig:comp_md_prism}
\end{figure}

Figure~\ref{fig:comp_md_prism} shows how the spinodal line as determined from the PRISM integral equation theory\cite{sung2005integral} depends on the parameter $\lambda$ (open symbols). As usual for demixing phenomena of polymeric systems, the relevant parameter is $4f_B(1-f_B)\beta_s \varepsilon$, where $\beta_s$ is the inverse of the spinodal temperature and $\varepsilon=-(\varepsilon_{AA}+\varepsilon_{BB}-2\varepsilon_{AB})/2$. (Note that in our case $\varepsilon > 0$.) The two theoretical curves that are shown in the figure correspond to $f_B=0.1$ and $f_B=0.5$, i.e.~to a weak concentration of the B monomers and to the case of a symmetric mixture. A comparison of these two curves shows thus that $4f_B(1-f_B)\beta_s \varepsilon$ does not only depend on $\lambda$, but also on $f_B$, but that this latter dependence is relatively weak.

Also included in the figure are the results from our simulations for different values of $\rho_m$ and $T$ (full symbols, right scale of the ordinate). Note that to obtain these points we have followed the system at a given $T$ and for each time we have measured the value of $f_B$ and $\lambda$ and thus drawn a point in the phase diagram showing $4f_B(1-f_B)\beta \varepsilon$ (where $\beta$ is the inverse temperature) as a function of $\lambda$ (Fig.~\ref{fig:comp_md_prism}, right scale). The system is initially in the homogeneous region of the phase diagram and with the progression of the catalytic process moves in the plane spanned by the parameters $\lambda$ and $4f_B(1-f_B)\beta \varepsilon$. The figure shows that this trajectory is basically independent of temperature or the density of the monomers, indicating that from this point of view the process is universal.  In addition we see that at short and intermediate times the data are basically falling on a vertical line that passes through the point $\lambda=0$ which implies that for these times the polymers are indeed random. Only for very long times the data points bend to the left, i.e.~$\lambda<0$, indicating that the internal structure of the polymer chains has homogeneous blocks that are on average smaller than the ones expected for a purely random chain with the same value of $f_B$, in agreement with the data presented in Fig.~\ref{fig:Sq_reac_vs_T_v2}.

Although the different combinations of $\rho_m$ and $T$ give rise to the same trajectory in the plane spanned by $\lambda$ and $4f_B(1-f_B)\beta_s \varepsilon$, the time at which the system enters into the theoretical spinodal region does depend on $\rho_m$ and $T$.  We find that this crossing occurs at $t_B^x \approx 120$ for $\rho_m=0.3$ and $T=0.25$, at $t_B^x \approx 40$ for $\rho_m=0.6$ and $T=0.35$, and at $t_B^x \approx 15$ for $\rho_m=0.6$ and $T=0.25$. From Fig.~\ref{fig:max_Sbq_tb}a we recognize that this crossing time corresponds to the time at which $\max(S_B(q))$ starts to show a plateau. This is thus evidence that the crossing of the spinodal line in Fig.~\ref{fig:comp_md_prism} corresponds to the final stage of the growth of the clusters. Hence we can identify the time $t_B$ at which $S_B(q)$ starts to grow as the time at which the system crosses the binodal (see Fig.~\ref{fig:phase_diagram}) and thus starts to enter a metastable phase that is characterized by the presence of clusters. With progressing time the system enters into the theoretical spinodal region and leaves this metastable phase via a spinodal decomposition at the time on the order of $t_B^x$.

Our simulations show that once the system crosses the critical line it becomes structured in the form of a cluster phase, see Fig.~\ref{fig:max_Sbq_ts}a. This type of micro-separated mesophase is not the one foreseen by the analytical theories which have instead predicted a macrophase separation if $\lambda$ is around zero~\cite{sung2005integral}. One possible explanation for this discrepancy might be the fact that in our system we are looking at an out-of-equilibrium process and therefore the resulting structure cannot be mapped in a simple manner onto the one of an equilibrium system, even if the internal structure of the chains are the same. On the other hand also Slimani {\it et al,} found in their simulation of a random copolymer melt that the system does make a microphase separation~\cite{slimani2013Macromol}. Thus the out-of-equilibrium situation might not really be the reason for the difference between theory and simulations and in the following we will give evidence that the explanation lies probably elsewhere. More studies on this are thus needed to clarify this point. 

\begin{figure}
\includegraphics[width=\linewidth]
{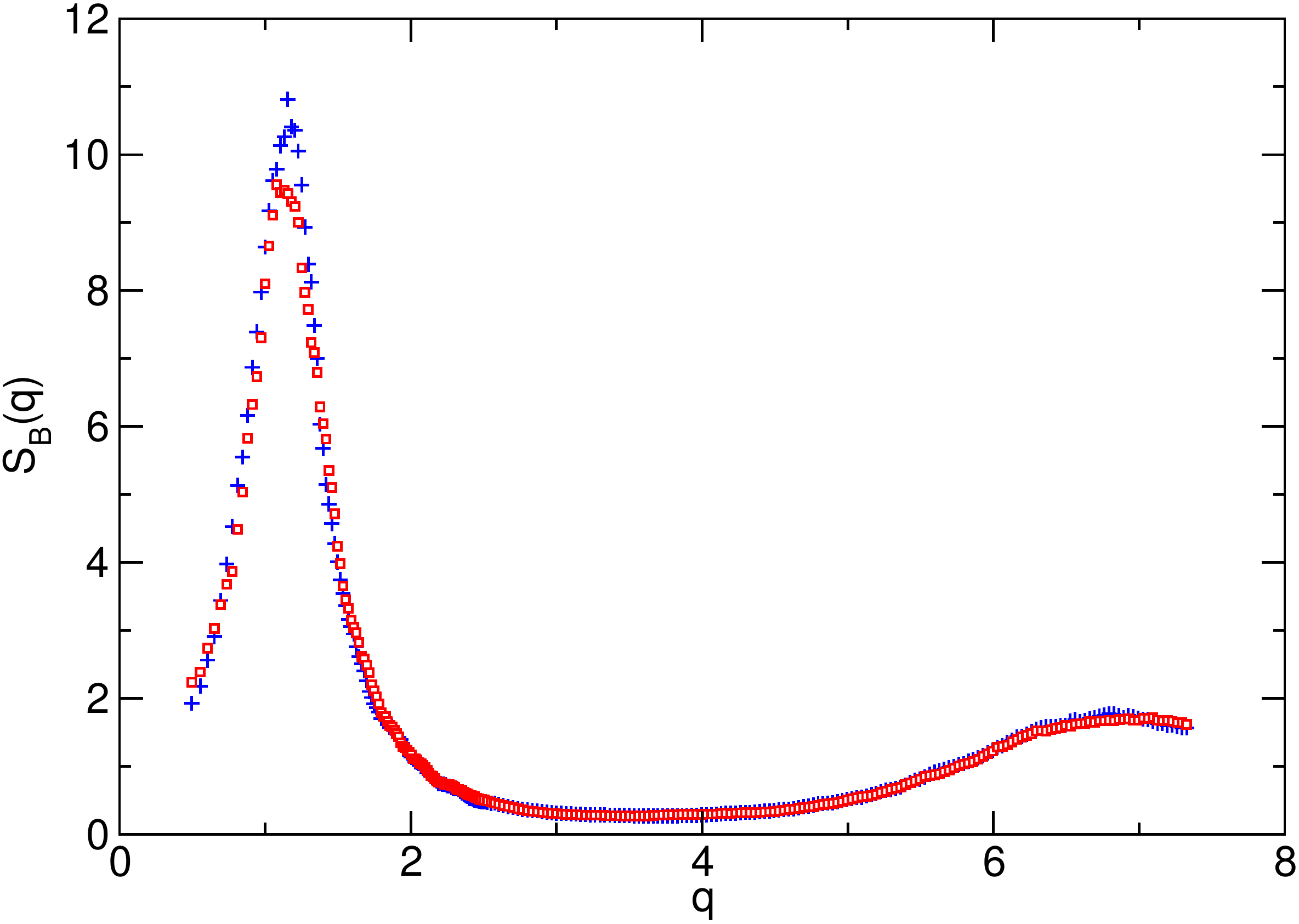}
\caption{(Colour online) Partial structure factors of the B monomers $S_B(q)$. Blue symbols: For a catalytically-induced structure of the polymers at $f_B=0.2$, $T=0.25$ and $\rho_m=0.6$; Red symbols: For the same set of copolymers (same A-B sequences), starting from $T=1.0$ and quenching the system to $T=0.25$. $S_B(q)$ is computed after $8.10^6$ steps at $T=0.25$.}
\label{fig:Sq_reac_vs_T}
\end{figure}

To investigate whether or not the cluster phase is related to the out-of-equilibrium dynamics of the system, we have taken a sample that showed a clear cluster phase ($T=0.25$ for $f_B=0.2$, $\rho_m=0.6$) and have frozen in the internal structure of the chains, i.e.~we stopped the catalytic reaction. This system of random copolymers was then heated to high temperature, $T=1.0$, and equilibrated at this $T$. Subsequently we have cooled down this system back to $T=0.25$ and used $10^7$ time steps to reach the equilibrium state. In Fig.~\ref{fig:Sq_reac_vs_T} we show the partial structure factor $S_B(q)$ for the two systems, i.e.~the one with the catalytically-induced structure at $f_B=0.2$, $T=0.25$ and $\rho_m=0.6$ and the one of the system with the same set of copolymers that was heated and cooled. We notice that the peak is about $10\%$ higher in the catalytically-induced structure and its width is slightly enhanced at low $q$ in the heated/cooled structure. Hence we can conclude that the structure of the system depends slightly on its history and not only on the chemical composition of the chains.

The analytical results on the random copolymers allow to test whether it is possible to obtain a semi-quantitative understanding of some of the results that we got from our simulations. To this aim we return to Fig.~\ref{fig:comp_md_prism} from which we have concluded that the phase separation occurs if $4 f_B (1-f_B) \beta\varepsilon$ reaches a critical value $\chi_c$, i.e.

\begin{equation}
\chi_c = 4 f_B (1-f_B) \beta\varepsilon \quad .
\label{eq1}
\end{equation}

Note that in our case the temperature is kept fixed and that instead $f_B$ is the variable. It is obvious that in Eq.~(\ref{eq1}) we can replace $f_B$ by $f_A=1-f_B$ without changing the contents of the equation and thus we will replace $f_B$ by $f$. The solutions of this equation are given by 

\begin{equation}
f_c = \frac{1}{2}\left(1\pm \sqrt{1- \chi_c T / \varepsilon}\right) \quad .
\label{eq2}
\end{equation}

\noindent
The relevant solution for our case is the smaller one, i.e.~the one with the minus sign. Since in our case $\lambda$ is close to zero the critical value $\chi_c$ is around 0.1 (see Fig.~\ref{fig:comp_md_prism}). Since here the value of $\varepsilon$ is 0.5 one can make a Taylor expansion of the right hand side of~Eq.(\ref{eq2}) which gives

\begin{equation}
f_c = \frac{\chi_c T}{4 \varepsilon} \quad .
\label{eq3}
\end{equation}

From the inset of Fig.~\ref{fig:nb_B_arrhenius}c we have:

\begin{equation}
f=r_0 \exp(-\Delta/T) t
\label{eq4}
\end{equation}

\noindent
where $r_0$ is a constant and $\Delta=-2.94$, see Fig.~\ref{fig:nb_B_arrhenius}c. If we denote by $t_c$ the time it takes to reach the critical concentration $f_c$ we thus have

\begin{equation}
f_c(T)=r_0 \exp(-\Delta/T) t_c \quad .
\label{eq5}
\end{equation}

Equating Eqs.~(\ref{eq3}) and (\ref{eq5}) gives thus

\begin{equation}
t_c(T) = \frac{\chi_c}{4 \varepsilon r_0} T \exp\left(\frac{\Delta}{T}\right) \quad .
\label{eq6}
\end{equation}

Expressing this time in the time scale $t_B\propto \exp(-\Delta/T)$ gives thus 

\begin{equation}
t_B^c(T)= A T \quad ,
\label{eq7}
\end{equation}

\noindent
where $A$ is a constant. Thus we conclude that mean field theory, which is the basis of Eq.~(\ref{eq1}), predicts that the time at which the system reaches the point at which it becomes thermodynamically unstable is directly proportional to temperature. In the context of Fig.~\ref{fig:comp_md_prism} we have estimated the time $t_B^x$ at which the simulated system crosses the theoretical spinodal boundary and found  $t_B^x(T=0.35)=40$ and $t_B^x(T=0.25)=15$. The simulation gives thus a ratio between the two times of 2.7 whereas the expression (\ref{eq7}) predicts 1.4. This discrepancy is not very surprising since, as mentioned above, the expression (\ref{eq4}) gives a good description of the time-dependence of $f$, but does not contain any information on the dynamical reaction of the system due to the presence of the attractive B monomers. (This latter information is encoded in the time scale $t_S$.) If we return to the schematic Fig.~\ref{fig:phase_diagram} we thus can say that the mean field theory is able to describe the static properties of the system between $\rho_B=0$ and the binodal line (full line) as well as the location of the spinodal line (dotted line). The dynamics characterizing the restructuring of the system (growth of the peak in $S(q)$ between the binodal and the spinodal line, and the subsequent coarsening of the system at long times) is, however, not accessible to this theoretical prediction.

\section*{Conclusions}

Our simulations of this catalyst-induced gelation process show that the resulting polymer gels can have a microscopic quasi-ordered cluster phase, i.e.~a structure that has not been previously observed in gels that are formed by a quench in temperature or by the addition of a chemical agent. We have found that certain details of this process, such as its $T-$dependence, can be understood in a simple manner from the effective reaction rate of the catalyst and the strength of the attraction between the monomers. The internal structure of the chains, i.e.~the sequence of the A and B monomers, resulting from the catalytic reactions can be discussed in the context of random copolymers with chemical correlations and we show that theoretical results~\cite{sung2005integral} and simulations are consistent with regards to the location of the homogeneous and heterogeneous regions. The theoretical calculations were so far not able to predict the nature of the unstable phase and therefore our results, showing that this phase is given by liquid-like clusters, is an interesting extension of these calculations. Finally we show that the structuring observed for a set of copolymers generated via catalytic reactions is slightly different from that obtained by a temperature quench of the same set of copolymers, demonstrating that the details of the cluster phase are the result of an out-of-equilibrium process. This understanding will thus allow to produce also in real life gels with such ordered microstructure and hence this catalytic reaction is a new approach to design materials with novel structures and mechanical properties as they are needed, e.g., for scaffolds in tissue engineering~\cite{drury2003}.

\section*{Acknowledgements}
The authors thank J.-L. Barrat, J. Baschnagel, C. Garnier, J. Oberdisse, and A. Yethiraj for useful discussions. W.K. is member of the Institut Universitaire de France. This work has been supported by LabEx NUMEV (ANR-10-LABX-20) funded by the ``Investissements d'Avenir'' French Government program, managed by the French National Research Agency (ANR). Simulations were performed at the Center of High Performance Computing HPC@LR in Montpellier.

\balance

\bibliographystyle{unsrt} 
\bibliography{enzyme} 

\end{document}